\documentclass[aip,reprint]{revtex4-1}
\usepackage{graphicx}
\usepackage{float}
\draft 

\begin{document}


\title{Cation dependent electroosmotic flow in glass nanopores} 



\author{Jeffrey Mc Hugh}
 \affiliation{Cavendish Laboratory, University of Cambridge, Cambridge CB3 0HE, United Kingdom}
\author{Kurt Andresen}
 \affiliation{Gettysburg Department of Physics, Gettysburg College, Gettysburg, PA 17325, United States}
\author{Ulrich F. Keyser}
 \email[Email: ]{ufk20@cam.ac.uk}
\affiliation{Cavendish Laboratory, University of Cambridge, Cambridge CB3 0HE, United Kingdom}


\date{\today}

\begin{abstract}
We present our findings on the changes to electroosmotic flow outside glass nanopores with respect to the choice of Group 1 cation species. In contrast with standard electrokinetic theory, flow reversal was observed for all salts under a negative driving voltage. Moving down Group 1 resulted in weaker flow when the driving voltage was negative, in line with the reduction in the zeta potential on the glass surface going down the periodic table. No trend emerged with a positive driving voltage, however for Cs, flow was uniquely found to be in reverse. These results are explained by the interplay between the flow inside the nanopore and flow along the outer walls in the vicinity of the nanopore.
\end{abstract}


\maketitle 


Nanopores are sensors based on the resistive-pulse technique\cite{Bayley2000}. Sensing is achieved by monitoring the ionic current through a nanoscale aperture in electrolytic solution. Nanopores exist in a variety of forms, the earliest used for sensing being the biological nanopore $\alpha$-haemolysin\cite{Division1996, Jones1996}. Today many solid state nanopore systems are known, primary examples of these being Si$_3$N$_4$\cite{Keyser2006}, quartz glass\cite{Steinbock2010} and graphene\cite{Merchant2010}. They have all proven capable of single molecule sensing\cite{Howorka2009}, detecting proteins\cite{Han2006, Plesa2013}, DNA sequencing\cite{Schneider2012a} and, in conjunction with DNA nanotechnology, detection of single nucleotide polymorphisms\cite{Kong2017} and specific proteins from mixtures\cite{Bell2016}.

\begin{figure*}[!htbp]
\includegraphics[width=0.8\textwidth]{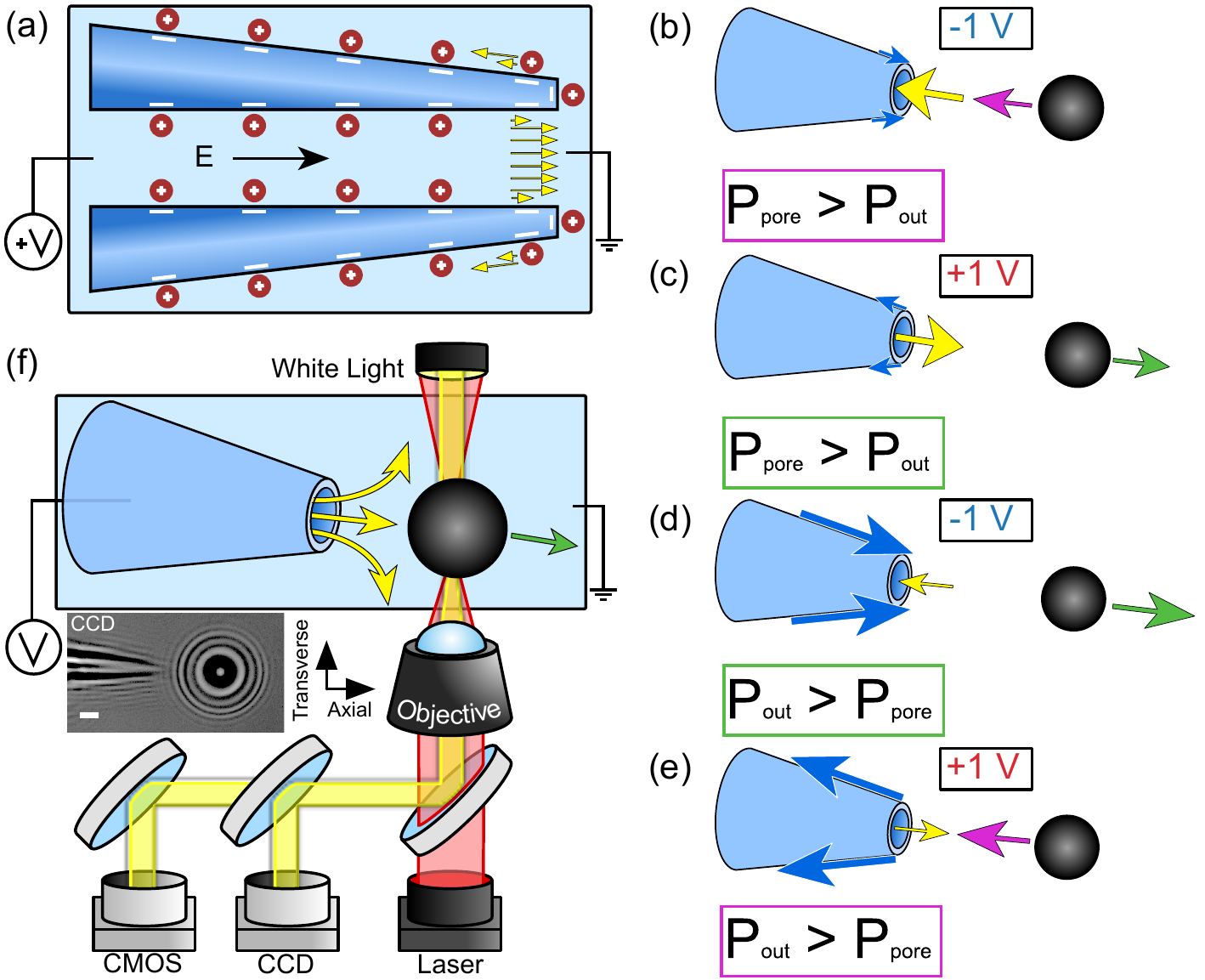}
\caption{\label{fig1_cap} Probing glass nanopore EOF using optical tweezers, and sketches of expected net outcomes.
(a) The nanopore surface is negative, leading to positive charge accumulation locally. The charges move under the applied field establishing flow.
(b) Here yellow arrows represent flow through the nanopore, blue arrows represent flow along the nanopore walls, purple arrows represent a resultant force directed towards the nanopore and green arrows represent a resultant force away from the nanopore. Arrow sizes represent the relative magnitudes of the flow. At $-1$ V, the inflow at the pore and $P_{pore}$, are stronger than the flow along the outer walls and $P_{out}$, resulting in a flow field directed towards the pore.
(c) At $+1$ V $P_{pore}$ is stronger than $P_{out}$ resulting in a flow field directed away from the pore.
(d) At $-1$ V $P_{pore}$ is weaker than $P_{out}$ resulting in a flow field directed away from the pore.
(e) At $+1$ V $P_{pore}$ is weaker than $P_{out}$ resulting in a flow field directed towards the pore.
(f) The setup consists of an IR laser focused through a microscope objective trapping a polystyrene bead positioned close to the nanopore. An applied voltage induces electroosmotic flow, displacing the bead. CCD and CMOS cameras track this displacement, facilitating determination of the flow forces. (Inset) Image of a capillary nanopore and a trapped bead viewed with the CCD camera, coordinate system for the experiment is shown on the right. Scale bar is $4$ $\mu$m.}
\end{figure*}

Hydrodynamic and electrokinetic phenomena dictate the behavior of analytes in nanopores. There are many works theoretically and experimentally probing the details of these phenomena with regards to micro- and nanofluidic systems\cite{Dutta2001, Rezaei2015, Nam2015, Monteferrante2015}. Here, of prime importance is electroosmosis. Si$_3$N$_4$ and glass nanopores have a negative surface charge in solution at biological pH. This results in a build-up of positive ions proximate to the surface\cite{Santiago2001}. Applying an electric field to drive an analyte through a nanopore causes the charges at the surface to move. The moving charges couple to the fluid medium and result in electroosmotic flow (EOF). This effect is depicted in Fig. \ref{fig1_cap}(a). The force a target molecule experiences in nanopores thus depends sensitively on the direction and strength of EOF\cite{VanDorp2009, Reiner2010}; it may slow the target down in a manner useful for sensing, or it may deny entry to molecules, hampering throughput\cite{Ghosal2007, Ghosal2007a, VanDorp2009}. As such, EOF in nanopores has been extensively studied\cite{Firnkes2010, Boukhet2016, Ghosal2019}, including reports of enhancement of molecular binding within an $\alpha$-haemolysin nanopore with EOF\cite{Gu2003}, facilitated protein capture in Fragaceatoxin C nanopores using EOF\cite{Huang2017}, and recently the demonstration that EOF can be used to control the folding state of DNA entering glass nanopores\cite{Ermann2018}.

Applying an electric field through the nanopore not only drives flow from within the pore, it establishes a flow field in the region outside the pore that is several microns in extent. This field can be quantified by a single parameter, $P$, the force required to generate this field in an otherwise calm fluid. This force originates from an immersed fluid jet which is described by the Landau-Squire solution to the Navier-Stokes equation\cite{Laohakunakorn2013} and is effective in describing the flow behavior resulting from the applied electric field. Laohakunakorn \textit{et al}\cite{Laohakunakorn2013, Laohakunakorn2015b} previously showed how geometry and concentration of K ions influence the magnitude and direction of this jet in glass nanopore systems; however the effect of salt choice has not been studied so far.

Key to understanding the flow behavior is the fact that voltage-induced flow is driven not only along the inside wall of the nanopore, but on the outer wall too. $P$ can be divided into two components, the force along the outer walls, $P_{out}$, and the force through the nanopore, $P_{pore}$, with the outer and inner walls having the same electrical double layer structure. The electric field applied through the nanopore results in the pore acting like a point charge\cite{Laohakunakorn2015b}, and there exists a small, finite electric field along the outer walls directed opposite to the field within the pore. $P_{pore}$ and $P_{out}$ therefore, must oppose each other because of their antagonistic driving fields. For a given electric field, $P_{pore}$ is limited by the no-slip boundary condition and hence, the area of the nanopore. However $P_{out}$ can grow ever larger as the only relevant boundary is the nanopore surface. Though the driving field is weaker than inside the pore, the area is effectively the extent of the fluid bath beyond the nanopore. While the outer flow will decay at large distances due to inertial effects, it still becomes the dominant contributor to the flow field, and thus the jet behavior.

With this in mind, Figs. \ref{fig1_cap}(b)--(e) depict the four net outcomes that result from the difference in magnitude and direction of $P_{pore}$ and $P_{out}$ under negative and positive applied voltages. In Fig. \ref{fig1_cap}(b) with a negative voltage applied, $P_{pore}$ is larger than $P_{out}$, and flow is thus directed towards the nanopore. In Fig. \ref{fig1_cap}(c) with a positive voltage applied, $P_{pore}$ is larger than $P_{out}$, leading to flow away from the nanopore. Figs. \ref{fig1_cap}(b) and (c) are the intuitive outcomes in a system with no outer wall. In Fig. \ref{fig1_cap}(d) with a negative voltage applied, $P_{pore}$ is smaller than $P_{out}$ and flow is thus directed away from the nanopore. In Fig. \ref{fig1_cap}(e) with a positive voltage applied, $P_{pore}$ is smaller than $P_{out}$, resulting in flow towards the nanopore. Figs. \ref{fig1_cap}(d) and (e) demonstrate the importance of the outer flow, $P_{out}$, in a confined system and illustrate flow reversal\cite{Laohakunakorn2015b}.

Despite ions of different elements having different sizes, the standard mean field theories that describe EOF as a surface derived effect do not account for salt species and their resultant differing surface charge densities. In this paper we utilize a highly sensitive apparatus which combines optical tweezers with nanopores to investigate electroosmotic flow in glass nanopores. Detecting down to sub-pN forces more than $5$ $\mu$m away from the pore (see Fig. S1), we quantified the voltage-induced flow fields about glass nanopores and demonstrate the importance of salt species as a parameter in nanofluidic systems.

By monitoring the force experienced by a trapped bead close to the nanopore while a voltage is applied, the details of the flow field can be extracted. A full description of the optical tweezers used in this experiment was previously published\cite{Otto2011b}.
Fig. \ref{fig1_cap}(f) shows the essentials of the experimental setup. It consists of a $1064$ nm ytterbium fiber laser focused through an inverted microscope objective ($60\times$ UPlanSApo water immersion, NA $1.2$, Olympus, Japan). Streptavidin coated polystyrene beads ($2$ $\mu$m diameter, Kisker, Germany) were suspended in salt solution in a PDMS walled bath with a glass coverslip base over the objective. The bath is illuminated directly from above with a white light source (DC-950 Fiber-Lite, Edmund Optics, USA).
 A bead trapped in the bath was monitored using a CCD camera (DMK31AF03, Imaging Source, Germany) and a high-speed CMOS camera (MC1362, Mikotron, Germany). Inset in Fig. \ref{fig1_cap}(f) is a cropped portion of the view from the CCD.
Using these two cameras, the position of a trapped bead can be monitored in three dimensions. The CMOS camera tracks the in-plane movement of the bead live at $1$ kHz with sub pixel accuracy using a previously described autocorrelation method\cite{Gosse2002}, leading to force measurements with sub-pN resolution. Force is ascertained from the spring constant, $k$, of the optical trap. $k$ is determined for in-plane motion by fitting a Lorentzian function to the power spectral density of the stochastic motion of a trapped bead\cite{Gittes1997}.

\begin{figure}[!htbp]
\includegraphics[width=0.48\textwidth]{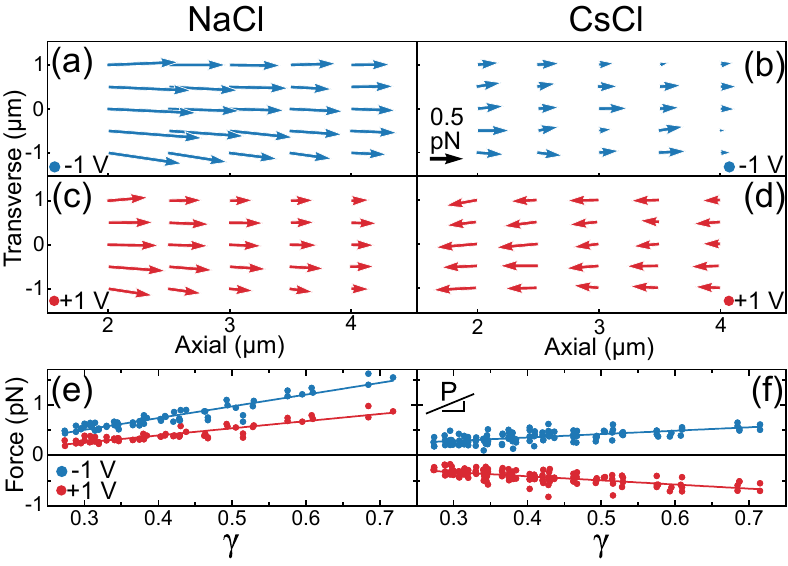}
\caption{\label{fig2_cap} Maps of flow forces around nanopores. (a) Map of flow forces measured when $-1$ V was applied in $10$ mM NaCl. (b) Map of flow forces measured when $-1$ V was applied in $10$ mM CsCl. Scale arrow is $0.5$ pN. (c) Map of flow forces measured when $+1$ V was applied in $10$ mM NaCl. The nanopore aperture was located at ($0$, $0$) in the transverse and axial directions. In (a)--(c) forces are directed away from the nanopores. (d) Map of flow forces measured when $+1$ V was applied in $10$ mM CsCl, these are directed towards the nanopore. (e) Plot of force data from (a) and (c) against the position parameter $\gamma$. (f) Plot of force data from (b) and (d) against $\gamma$. The linear fits follow the Landau-Squire jet model and yield the nanojet force $P$.}
\end{figure}

The nanopores are fabricated from quartz glass capillaries (Intracel, UK) pulled with a laser pipette puller (P-2000, Sutter Instruments, CA, USA). The nanopores have a diameter of approximately $150$~nm. The nanopores are mounted into a capillary holder which is plugged into the headstage of the electrophysiology amplifier (Axopatch 200b, Axon Instruments, USA). Prior to mounting the nanopore, it is plasma cleaned (Femto, Diener, Germany) and then immersed in the chosen salt solution in order for it to completely fill with solution. The headstage is mounted onto a micromanipulator (Patchstar, Scientifica, UK) allowing for programming the motion of the nanopore through a series of locations in a plane relative to the trapped bead. Control and automation of the setup are achieved through custom LabVIEW code (LabVIEW 2016, National Instruments).
At each location the deflection of the bead was recorded while the voltage was varied from $0$ V, to $-1$ V, to $+1$ V, and back to $0$ V. Voltage was applied using Ag/AgCl electrodes, with an electrode inside the capillary holder and the ground electrode located in the bath. The random deflection of the bead at $0$ V was used to ensure the only force measured was that due to EOF.
For each measurement a bead was trapped with stiffness, $k= 0.03$ pN nm$^{-1}$. The measurement plane had two directions, the axial direction parallel to the long axis of the glass capillary and the transverse direction perpendicular to that. These are shown in Fig. \ref{fig1_cap}(f). Flow measurements were conducted starting with the bead centre $2$ $\mu$m axially from the nanopore. This position was chosen as optimum, minimizing the risk of the bead being driven from the trap by very strong flows while still probing close to the nanopore aperture.
For each position in the measurement plane, the mean deflections of the trapped bead at $0$ V, $-1$ V and $+1$ V were calculated. These were converted to forces in the manner described above, with the force at $0$ V subtracted from that at $-1$ V and $+1$ V. Nanopore geometry and salt concentration were kept consistent across experiments, while the salt choice was varied through the range of Group 1 chlorides. Flow fields were recorded for each salt with multiple nanopores to assess variability.

Fig. \ref{fig2_cap} consists of four different maps of the flow force observed over an array of positions relative to the nanopore. In Figs. \ref{fig2_cap}(a) and (b) $-1$ V was applied and fluid flow in NaCl and CsCl solution was directed away from the nanopore. Intuitively, the strength of the flow forces depend on distance from the nanopore, with the highest forces generally along the line of the central axis of the capillary, and in close vicinity to the nanopore. Forces were higher with NaCl than CsCl.
Fig. \ref{fig2_cap}(c) is a typical flow field observed in NaCl when $+1$ V was applied. The forces are directed away from the nanopore and again depend on distance from the nanopore. Outflow in this situation is the intuitively expected result and is in agreement with previous reports for KCl at this concentration\cite{Laohakunakorn2015b}.
Fig. \ref{fig2_cap}(d) is a flow field recorded in CsCl with $+1$ V applied. Flow was observed to be directed back towards the nanopore, with the greatest forces still observed closest to the pore. This flow reversal at $10$ mM concentration was unexpected.
We can quantify these flow fields using the Landau-Squire solution\cite{Laohakunakorn2013}, which allows us to linearize the force data and determine $P$, the nanojet force which generated the flow field. To achieve this the position coordinates of each force measurement were first transformed into a single position parameter, $\gamma$ (see SI Section 4). Fig. \ref{fig2_cap}(e) is a plot of the NaCl force data from Figs. \ref{fig2_cap}(a) and (c) plotted as a function of $\gamma$. The slope of the linear fit gives the parameter $P$ for the salt at each applied voltage, and also demonstrates that the Landau-Squire solution is a good model for the fluid flow observed. Fig. \ref{fig2_cap}(f) is the same plot but with the CsCl force data from Figs. \ref{fig2_cap}(b) and (d). Note the flow reversal which is evident from the negative slope for the positive voltage force data with Cs.

\begin{figure}[!htbp]
	\includegraphics[width=0.38\textwidth]{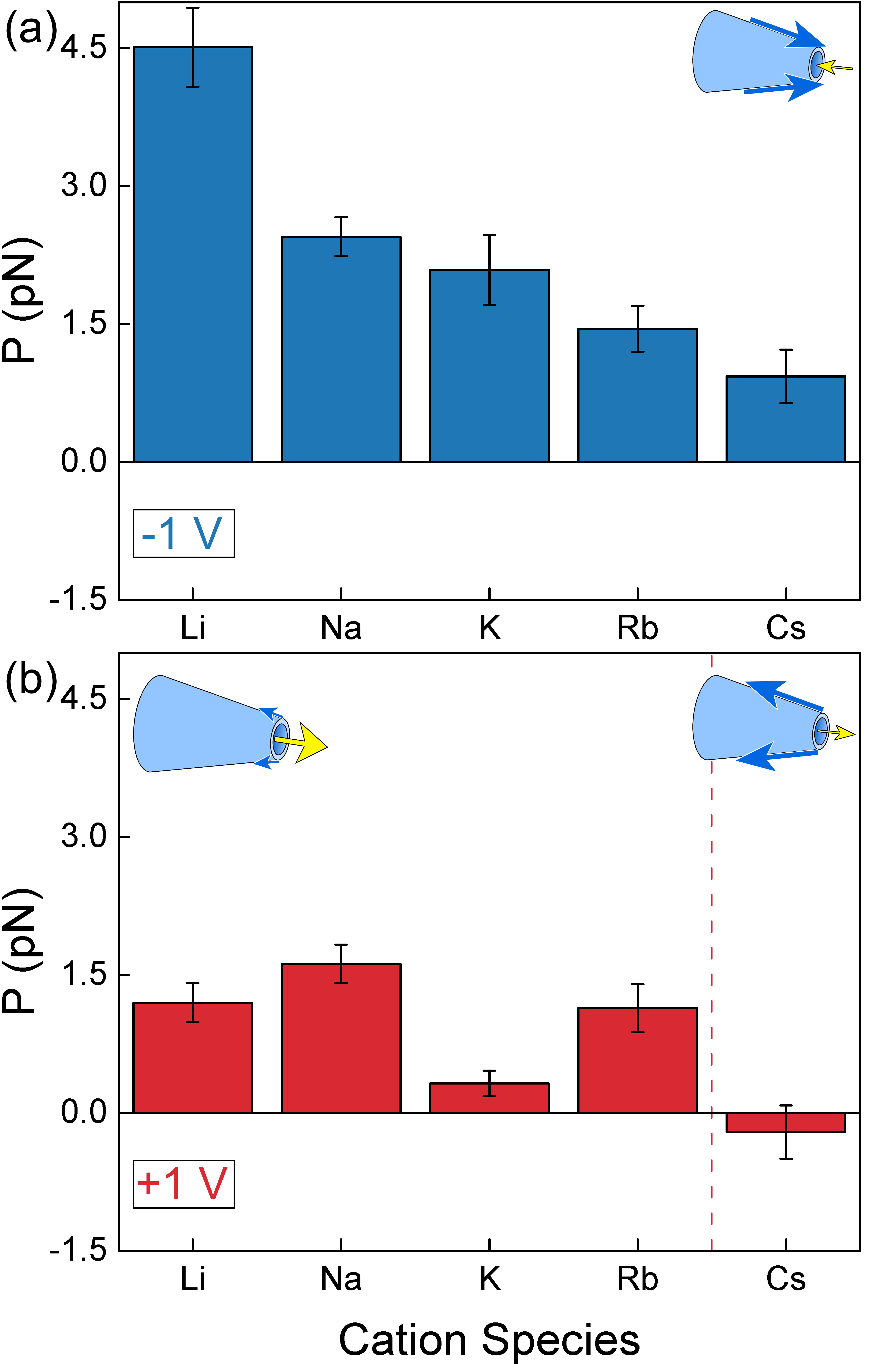}
\caption{\label{fig4_cap} Cation dependent variations in nanopore EOF. (a) $P$ for the first 5 cations in Group 1 of the periodic table with $-1$ V applied. Positive values are flows away from the pore. (b) $P$ for the same cations with $+1$ V applied. Only CsCl shows a reversal in flow direction. Error bars are $95\%$ confidence intervals. Inset are the flow models that were observed.}
\end{figure}

The dramatic difference between Na and Cs demonstrates how $P$ is the result of an interplay between $P_{pore}$ and $P_{out}$. Given this result, to thoroughly understand the effect of salt choice on nanopore EOF, we measured and compared the behavior of all Group 1 chloride salts. $P$ was calculated from force maps recorded for each salt and Figs. \ref{fig4_cap}(a) and (b) are plots of these comparisons for $-1$ V and $+1$ V respectively. $95\%$ confidence intervals are shown for the $P$ value of the various salts, the size of these intervals is attributed predominantly to variability between nanopore geometries. Inset in Figs. \ref{fig4_cap}(a) and (b) are illustrations of the relative strengths of $P_{pore}$ and $P_{out}$ that produce these results.
In (a), where $-1$ V was applied for each salt, the flow was always positive, directed away from the nanopore. $P$ gets smaller with each cation as we move down Group 1, with the weakest flows measured for Cs.
In Fig. \ref{fig4_cap}(b), when the bias was positive, the first four salts of Group 1 behave in the same overall manner. The flow is positive but here there is no clear trend in the flow magnitude with cation choice. The net flow with Cs is negative, directed back towards the nanopore. Cs is highlighted for this unique result, demonstrating, alongside the rest of our results, the complex behavior of EOF in glass nanopores.

In conclusion a highly sensitive method for quantifying nanopore EOF has been demonstrated. It has been used to elucidate the effect of cation species on the hydrodynamic environment of nanopores, with salt choice changing both the strength and direction of the flow field about the nanopores. The inner and outer flow model presented accounts for the different flows observed. The variation due to different salt species can be attributed to changes in the electrical environment of the nanopore surface. Cations further down Group 1 of the periodic table were found to have decreasing values of $P$ and thus weaker flow about the nanopore. The zeta potential of these salts on quartz glass is the most negative for Li and is less negative with each cation down the periodic table\cite{Ma2005} (see Fig. S2). EOF velocity is linearly dependent on zeta potential, possibly explaining the weakening of the flow with negative voltage, but not the behavior seen with a positive bias.
Most of the salts were found to exhibit weaker flow for the positive bias than the negative. Flow was still directed away from the nanopore with the lone exception of Cs, where flow reversal was observed. The flow reversal observed with Cs is quite striking particularly given the similarity of Rb and Cs in terms of zeta potential and ionic mobility. Additionally, the flow reversal observed with CsCl was previously reported for KCl at a concentration $20$ times more dilute than here\cite{Laohakunakorn2015b}, and is explained by $P_{out}$ becoming the dominant flow as in Fig. \ref{fig1_cap}(e) when Cs is the cation used. Following on from the findings we present here, nanopore translocation experiments with Cs could yield better understanding and improved results.

Our results demonstrate that electrokinetic phenomena have a considerable impact on the fluid environment about a nanopore, with flow readily observable at $> 5$ $\mu$m away from the pore. We show the complex interplay of ion type and fluid behaviour which provides an opportunity for theorists to further inform models of EOF in nanofluidics. Greater understanding of the parameters dictating nanopore EOF helps the design of nanopore sensing systems while also offering an opportunity to investigate the combined roles of surface chemistry and fluid dynamics in a confined nanoscale system.
\newline

See supporting information for plots of $P$ against zeta potential, extended force maps, a description of position calibration and details of the linearization procedure to obtain $\gamma$.
\newline

The authors thank A. L. Thorneywork and J. Jadwiszczak for careful reading of the manuscript and useful discussions and advice. J. Mc H. acknowledges funding from AFOSR (Grant No. FA9550-17-1-0118). U. F. K. is supported by ERC Consolidator Grant DesignerPores 647144.



%
%

%




\bibliography{eofcatbib_1.bib}

\end{document}